\documentclass[aps,prb, reprint,superscriptaddress]{revtex4-1}
\usepackage{amsmath}
\usepackage{graphicx}
\usepackage{dcolumn}
\usepackage{bm}
\usepackage{textcomp}
\usepackage{float}
\usepackage{amssymb}
\usepackage[colorlinks=true, citecolor={blue}, urlcolor={blue}, linkcolor = {blue}]{hyperref}

\newcommand{\beq}{\begin{equation}}
\newcommand{\eeq}{\end{equation}}
\newcommand{\beqa}{\begin{eqnarray}}
\newcommand{\eeqa}{\end{eqnarray}}

\begin{document}

\title{Supercurrent mediated by helical edge modes in bilayer graphene}

\author{Prasanna Rout}
\affiliation{QuTech and Kavli Institute of Nanoscience, Delft University of Technology, 2600 GA Delft, The Netherlands}

\author{Nikos Papadopoulos}
\affiliation{QuTech and Kavli Institute of Nanoscience, Delft University of Technology, 2600 GA Delft, The Netherlands}

\author{Fernando Pe\~naranda}
\affiliation{Instituto de Ciencia de Materiales de Madrid (ICMM), CSIC. Sor Juana In\'{e}s de la Cruz 3, 28049 Madrid, Spain}

\author{Kenji Watanabe}
\affiliation {Research Center for Functional Materials,
National Institute for Materials Science, Tsukuba 305-0044,
Japan}

\author{Takashi Taniguchi}
\affiliation {International Center for Materials
Nanoarchitectonics, National Institute for Materials Science,
Tsukuba 305-0044, Japan}

\author{Elsa Prada}
\affiliation{Instituto de Ciencia de Materiales de Madrid (ICMM), CSIC. Sor Juana In\'{e}s de la Cruz 3, 28049 Madrid, Spain}

\author{Pablo San-Jose}
\affiliation{Instituto de Ciencia de Materiales de Madrid (ICMM), CSIC. Sor Juana In\'{e}s de la Cruz 3, 28049 Madrid, Spain}

\author{Srijit Goswami}
\affiliation{QuTech and Kavli Institute of Nanoscience, Delft University of Technology, 2600 GA Delft, The Netherlands}

\begin{abstract}
Bilayer graphene encapsulated in tungsten diselenide can host a weak topological phase with pairs of helical edge states. The electrical tunability of this phase makes it an ideal platform to investigate unique topological effects at zero magnetic field, such as topological superconductivity. Here we couple the helical edges of such a heterostructure to a superconductor. The inversion of the bulk gap accompanied by helical states near zero displacement field leads to the suppression of the critical current in a Josephson geometry. Using superconducting quantum interferometry we observe an even-odd effect in the Fraunhofer interference pattern within the inverted gap phase. We show theoretically that this effect is a direct consequence of the emergence of helical modes that connect the two edges of the sample. The absence of such an effect at high displacement field, as well as in bare bilayer graphene junctions, confirms this interpretation and demonstrates the topological nature of the inverted gap. Our results demonstrate the coupling of superconductivity to zero-field topological states in graphene.
\end{abstract}

\maketitle

Helical edge modes in two-dimensional (2D) systems are an important building block for many quantum technologies, such as dissipationless quantum spin transport\cite{murakami_2003_dissipationless,roth_2009_nonlocal,brune_2012_spin}, topological spintronics\cite{tokura_2019_magnetic,he_2022_topological}, and topological quantum computation\cite{prada_andreev_2020}. Electrons traveling along these modes cannot invert their propagation direction unless their spin is flipped. As a consequence, they cannot backscatter as long as time-reversal symmetry is preserved (e.g. even in the presence of arbitrary non-magnetic defects). Helical states are expected to appear at the edges of 2D topological insulators\cite{fu_superconducting_2008-2,Beenakker_fermion_2013} or in one-dimensional semiconductors with large spin-orbit coupling (SOC)\cite{prada_andreev_2020}. Interestingly, single-layer graphene was the first theoretically predicted quantum spin Hall insulator~\cite{kane_quantum_2005}, whereby the intrinsic SOC gives rise to helical edge states. However, the strength of this Kane-Mele type SOC ($\lambda_{\rm{KM}}$) is too small ($\approx$ 40 $\mu$eV) to realize a topological phase in practice.

\begin{figure}
  \begin{center}
    \includegraphics[width=\columnwidth]{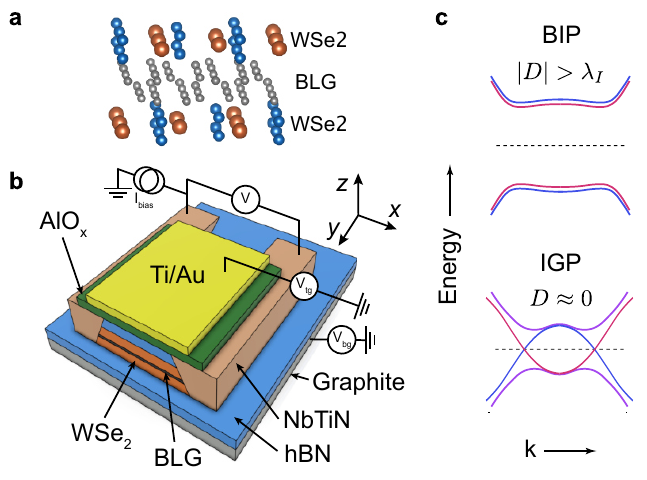}
  \end{center}
\caption{\textbf{Band inversion in a bilayer graphene junction.} \textbf{a} The BLG is symmetrically encapsulated in $\mathrm{WSe_2}$ leading to a proximity induced SOC. \textbf{b} Device schematic of a NbTiN-based Josephson junction (JJ). The hexagonal boron nitride (hBN) and hBN/AlO$_x$ act as bottom and top-gate dielectrics, respectively. The measurements are performed using a quasi-four terminal current-biased circuit. \textbf{c} Band structure of the encapsulated BLG for different displacement fields $D$. At high $|D|$ ($|u| > \lambda_{\rm{I}}$) the bulk is in a band-insulator phase (BIP), whereas the SOC creates an Ising gap in the bulk bands around $D = 0$, driving the system into an inverted-gap phase (IGP). In this weak topological phase, gapless edge states are present all around the BLG stack.}
\label{fig1}
\end{figure}

\begin{figure*}
  \begin{center}
    \includegraphics[width=\textwidth]{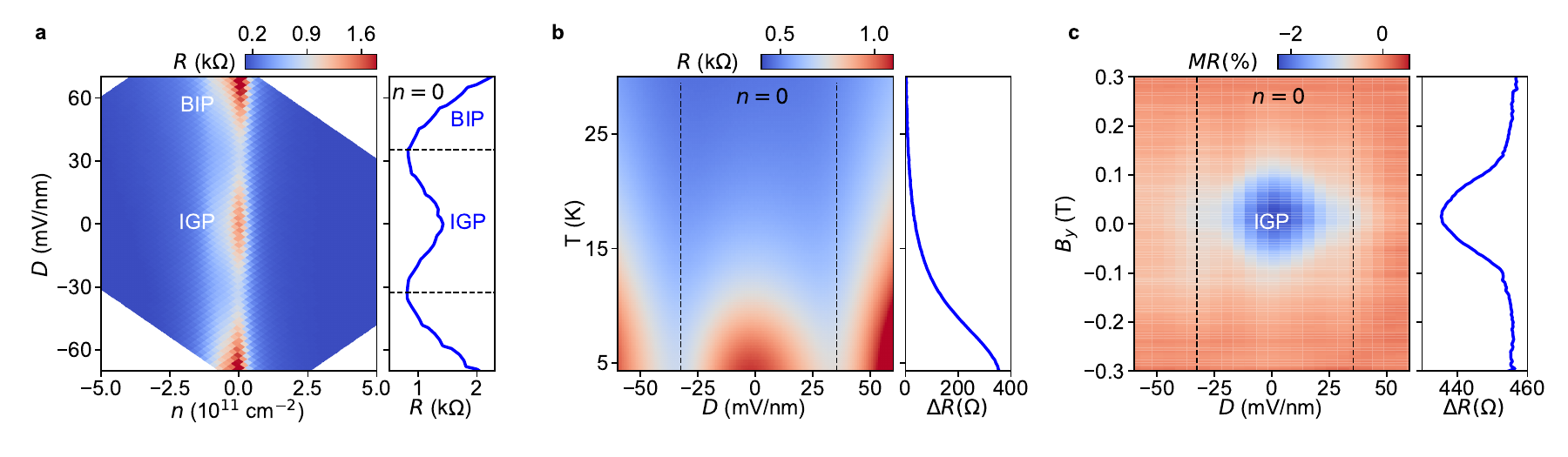}
  \end{center}
\caption{\textbf{Bulk and edge transport in the inverted-gap phase.} 
 \textbf{a} Resistance $R$ at 3.3 K for Dev~A measured as a function of carrier density $n = (C_{\rm{TG}}V_{\rm{TG}} + C_{\rm{BG}}V_{\rm{BG}})/e$ and displacement field $D = (C_{\rm{TG}}V_{\rm{TG}} - C_{\rm{BG}}V_{\rm{BG}})/2\epsilon_0$, where $C_{\rm{TG}(\rm{BG})}$ is the capacitance of top (bottom) gate. Side panel: line cut of the $R (n, D)$ map at $n = 0$. \textbf{b} Temperature dependence of $R$ measured at $n=0$ for Dev~A. Side panel: the disappearance of the IGP (between the two dotted lines) at higher temperatures is seen from the difference $\Delta R$ between $R$ at $D = 0$ and at the inversion point, $D_{\rm{inv}} = -32$ mV/mm. \textbf{c} The magnetoresistance $MR = [R(B_y) - R$(0.3 T)]$/R$(0.3 T) measured at 40 mK as a function of $D$ and in-plane magnetic field $B_y$ for Dev~A. The dotted lines represent the inversion points. The observed dip close to $D = 0$ and $B_y = 0$ is the result of the conduction due to helical edge channels. Side panel: field dependence of $\Delta R$.}
\label{fig2}
\end{figure*}

With the advent of graphene-based van der Waals heterostructures with exceptional electronic properties, several alternative approaches have been explored to create helical edge modes. These modes are shown to exist in the quantum Hall regime of single-layer graphene at filling factor $\nu = 0$ under the application of a large in-plane magnetic field~\cite{young_tunable_2014-1}, or using a substrate with an exceptionally large dielectric constant~\cite{veyrat_helical_2020-1}. Using a double-layer graphene heterostructure, helical transport was also observed by tuning each of the layers to $\nu = \pm{1}$~\cite{sanchez-yamagishi_helical_2017}. While these experiments did not involve superconductivity, they have been complemented by theoretical proposals showing that coupling the helical modes to a superconductor should give rise to topological superconductivity~\cite{san-jose_majorana_2015,Finocchiaro_2017_quantum}. Unlike the experiments involving topological insulators\cite{hart_induced_2014,pribiag_edge_2015,wiedenmann_4pi_2016,bocquillon_gapless_2017}, the main practical drawback of these proposals is the requirement of large magnetic fields, which is detrimental to any system involving superconductors.

Recently, it was shown that helical modes can appear at zero magnetic field in bilayer graphene (BLG) encapsulated with $\mathrm{WSe_2}$, a transition metal dichalcogenide (TMD)~\cite{island_spin-orbit_2019}. Several experiments have shown that graphene coupled to TMDs gives rise to a proximity induced Ising-type SOC, denoted $\lambda_{\rm{I}}$~\cite{wang_strong_2015-1,wang_origin_2016,wakamura_strong_2018,wakamura_spin-orbit_2019,zihlmann_large_2018,wang_quantum_2019,kedves_2023_stabilizing}. In the case of BLG symmetrically encapsulated in $\mathrm{WSe_2}$ (Fig.~\ref{fig1}a), $\lambda_{\rm{I}}$ has opposite signs on the two graphene layers, thereby effectively emulating a Kane-Mele type SOC, but with a significantly larger magnitude of few meVs\cite{zaletel_gate-tunable_2019,island_spin-orbit_2019, wang_quantum_2019}. Using an electric displacement field $D$ across the BLG, it was shown~\cite{island_spin-orbit_2019} that the bandstructure could be tuned continuously from a band-insulator phase (BIP) at large $D$, through a topological phase transition, and into an inverted-gap phase (IGP). 

In this work we combine such a $\mathrm{WSe_2}$/BLG/$\mathrm{WSe_2}$ heterostructure with a superconductor and study the Josephson effect across this phase transition. We show that the IGP is characterized by a local minimum of the critical current $I_c$. Supercurrents in this phase are expected to flow along the helical modes. We show theoretically that this should cause a unique even-odd modulation with flux in the Fraunhofer interference pattern of the Josephson junction (JJ), and present quantum interferometry measurements that clearly exhibit this signature. Importantly, the effect is absent in a trivial BLG system (i.e., without $\mathrm{WSe_2}$ encapsulation) as well as in the BIP of $\mathrm{WSe_2}$/BLG/$\mathrm{WSe_2}$, thus ruling out trivial edge transport as its origin, and strongly supporting the topological nature of the superconducting IGP.

A schematic of the devices is shown in Fig.~\ref{fig1}b (see Appendix A for details about the fabrication process). The $\mathrm{WSe_2}$/BLG/$\mathrm{WSe_2}$ heterostructure has superconducting NbTiN contacts as well as a back gate, $V_{\rm{BG}}$, and a top gate, $V_{\rm{TG}}$. We present results on two devices (Dev~A and Dev~B) fabricated on the same heterostructure. Dev~A has a larger contact separation, $L = $~3.7~$\mu$m, which suppresses the Josephson effect and allows us to study the normal state properties of the heterostructure. The supercurrent transport is explored in a ballistic JJ (Dev~B) with a contact separation of $L \approx $~300~nm. 

Figure~\ref{fig1}c shows how the bandstructure is predicted to evolve as a function of displacement field $D$. For $D=0$ the valence and conduction bands of the BLG are inverted due to the Ising SOC. The band inversion opens up a gap $\lambda_I$ in the bulk band structure. Increasing $|D|$ introduces an additional competing energy $u = - edD/\epsilon_{\rm{BLG}}$. Here, $d = 0.33$~nm is the BLG interlayer separation and $\epsilon_{\rm{BLG}} = 4.3$ is the out-of-plane dielectric constant of BLG. At large $|D|$, $u$ becomes the dominant energy scale compared to $\lambda_{\rm{I}}$ and a gap associated with layer-polarised bands opens up. Thus, by tuning $|D|$ one can transition from the inverted phase (IGP: $|u| < \lambda_{\rm{I}}$) to band-insulator phase (BIP: $|u| > \lambda_{\rm{I}}$) via an inversion point, where the bulk gap closes.  

Before proceeding to the superconducting regime of our devices, we characterize the various phases in the normal regime. As a first step we measure the dependence of normal resistance with bottom and top gates, which independently control the carrier density, $n$, and the displacement field, $D$. The resistance $R$ as a function of $n$ and $D$ reveals a local maximum in $R$ close to $D = 0$ along the $n = 0$ line. This is consistent with the predicted opening of a SOC gap in the IGP ($\lambda_{\rm{I}} > |u|$). A similar behavior has been observed previously in the form of an incompressible phase in capacitance measurements  \cite{island_spin-orbit_2019}. As we increase $|D|$, the gap is reduced and $R$ decreases, reaching a minimum at the gap-inversion point $\lambda_{\rm{I}} = |u|$. Increasing $|D|$ further leads to an increase in $R$ as the band-insulator gap grows. The inversion points corresponding to $R$ minima are at $D = -32$ and 35~mV/nm (Fig.~\ref{fig2}a), which yield $\lambda_{\rm{I}} = 2.5$ and $2.7$~meV.

The SOC gap can alternatively be probed by thermal excitation, see Fig.~\ref{fig2}b. With increasing temperature, $R$ decreases as expected for a thermally activated gap. In addition, the local $R$ maximum at $D = 0$ gradually becomes less pronounced as the temperature is increased, and completely vanishes at higher temperatures. To see this effect more clearly we determine the difference in the resistances at $D=0$ and at the inversion point $D_{\rm{inv}} = -32$~mV/mm, i. e., $\Delta R=R (D = 0) - R (D= D_{\rm{inv}})$. A positive value for $\Delta R$ implies the band inversion. This difference is almost zero at $T \approx 26$ K (side panel of Fig.~\ref{fig2}b), which corresponds to $\lambda_{\rm{I}} =$ 2.2~meV. The estimated $\lambda_{\rm{I}}$ matches quite well with previously reported values of $2.2-2.6$~meV \cite{island_spin-orbit_2019, wang_quantum_2019,kedves_2023_stabilizing}.

While the temperature dependence provides the information about the bulk gap, a verification of the existence and the helical character of edge states can be done by breaking time-reversal symmetry. An in-plane magnetic field $B_y$ introduces a Zeeman energy $E_{\rm{Z}}$ which opens a gap in the helical edge states \cite{island_spin-orbit_2019}, and removes their contribution from conduction. To check this effect we measure $R$ at $n = 0$ in the presence of an in-plane magnetic field ($B_y$) as presented in Fig.~\ref{fig2}c. At $D = 0$ we observed a dip in magnetoresistance $MR = [R(B_y) - R$(0.3 T)]$/R$(0.3 T) for $B_y <$ 0.15 T indicating the extra conduction from edge states. At higher fields the absence of these gapless states results in $MR$ values close to zero. Outside the IGP where no helical edge is present, $MR \approx 0$ for all $B_y$. Additionally, we can rule out that this effect is related to the bulk bands as $E_{\rm{Z}} < \lambda_{\rm{I}}$. Moreover, the positive $\Delta R$ for the complete field range is consistent with a band inversion originating from the SOC gap (side panel of Fig.~\ref{fig2}c). At lower fields we once again see the presence of conducting edge channels resulting in a dip in $\Delta R$.

\begin{figure}
  \begin{center}
    \includegraphics{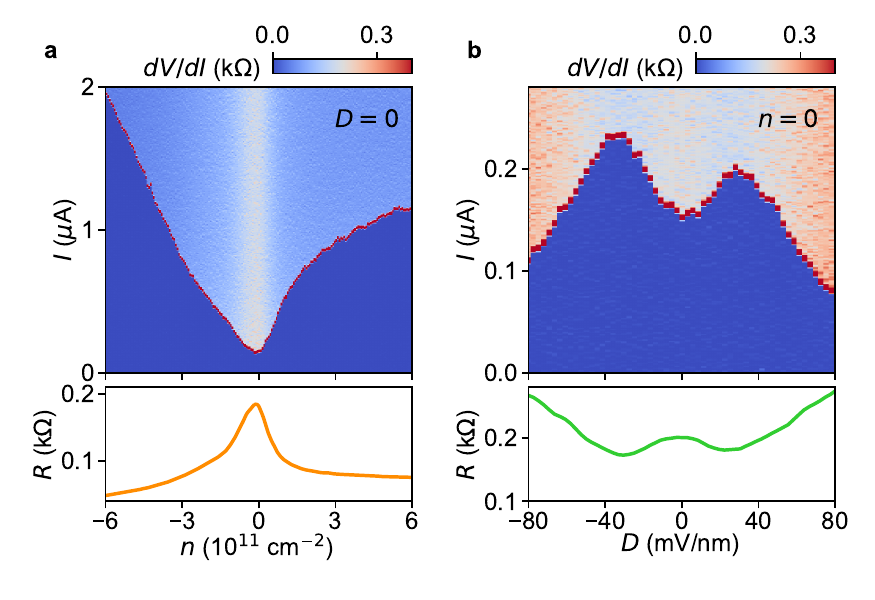}
  \end{center}
\caption{\textbf{Supercurrent in the inverted-gap phase.} \textbf{a} Current-bias dependence of the differential resistance $dV/dI$ for Dev B as a function of carrier density $n$ at $D = 0$ and $T =$~40 mK. The jump in $dV/dI$ marks the critical current $I_c$. The lower panel shows the normal-state resistance $R$($n$) for $D = 0$. \textbf{b} $dV/dI$ (upper panel) and normal-state resistance $R$ (lower panel) as a function of $D$ for $n = 0$.}
\label{fig3}
\end{figure}

After verifying the presence of helical edge states, we turn our focus towards inducing superconductivity in these edge modes using JJs. The JJ (Dev~B) is in the ballistic limit, as indicated by Fabry-Perot oscillations in the normal state transport~\cite{Calado:NN15,Ben-Shalom:NP16} (see Appendix C). We first check the evolution of the critical current ($I_c$) in the JJ as a function of $n$ for fixed $D=0$ (Fig.~\ref{fig3}a) and vice versa (Fig.~\ref{fig3}b). The dependence of $I_c$ on $n$ is similar to several previous studies \cite{Heersche:N07,Calado:NN15,Ben-Shalom:NP16,allen_2016_spatially,zhu_edge_2017}.
Figure~\ref{fig3}a shows an $I_c$ minimum at $n = 0$, coinciding with the $R$ maximum, while $I_c$ increases with higher electron/hole doping. The more interesting behavior is the $D$ dependence of $I_c$ at $n = 0$, displayed in Fig.~\ref{fig3}b. We observe a local minimum in $I_c$ at $D=0$, corresponding to the local $R$ maximum, and two maxima at higher $|D|$ values, concurrent to the $R$ minima at the gap inversion points. At higher $|D|$, $I_c$ decreases monotonically due to gradual opening of band-insulator gap \cite{zhu_edge_2017}.

The presence of a supercurrent and its suppression at $D=0$ is a signature of Josephson coupling across a BLG in an inverted phase. However, it does not reveal whether the supercurrent flows through the bulk or through edge channels. One could attribute the suppression of $I_c$ in the IGP to the opening of a bulk SOI gap entirely. Therefore, demonstrating the existence of proximitized edge modes requires a different probe that can differentiate edge from bulk supercurrents. A common method employed for this purpose is superconducting quantum interferometry (SQI)~\cite{hart_induced_2014,pribiag_edge_2015,allen_2016_spatially,bocquillon_gapless_2017,zhu_edge_2017,de_vries_he_2018,de_vries_crossed_2019}, which involves the measurement of $I_c$ as a function of a perpendicular magnetic field $B_z$. In the case of a JJ with homogeneous current transport across its width $W$, $I_c(B_z)$ should display a standard Fraunhofer pattern, following the functional form: $\sin(\pi\Phi/\Phi_0)/ (\pi\Phi/\Phi_0)$, where $\Phi = B_z/LW$ and $\Phi_0=h/2e$ is the superconducting flux quantum. This regime is clearly observed in our samples at high densities $n$, see Appendix E.
On the other hand, for a JJ with only two superconducting and equivalent edges (i. e., a symmetric SQUID) one expects a SQI pattern of the form $\cos(\pi\Phi/\Phi_0)$. However, the SQI pattern becomes more complicated if there is some coupling between these edges. An efficient inter-edge transport along edge channels flowing along the two SN interfaces (without becoming fully gapped by proximity) can lead to the electrons and holes flowing around the planar JJ, thereby picking up a phase from the magnetic flux. This phase introduces a $2\Phi_0$-periodic component into the $\Phi_0$-periodic SQUID pattern of the edge supercurrent~\cite{Baxevanis_even_2015}, which is a highly significant signature of the existence of proximitized helical modes.

\begin{figure*}
  \begin{center}
    \includegraphics[width=\textwidth]{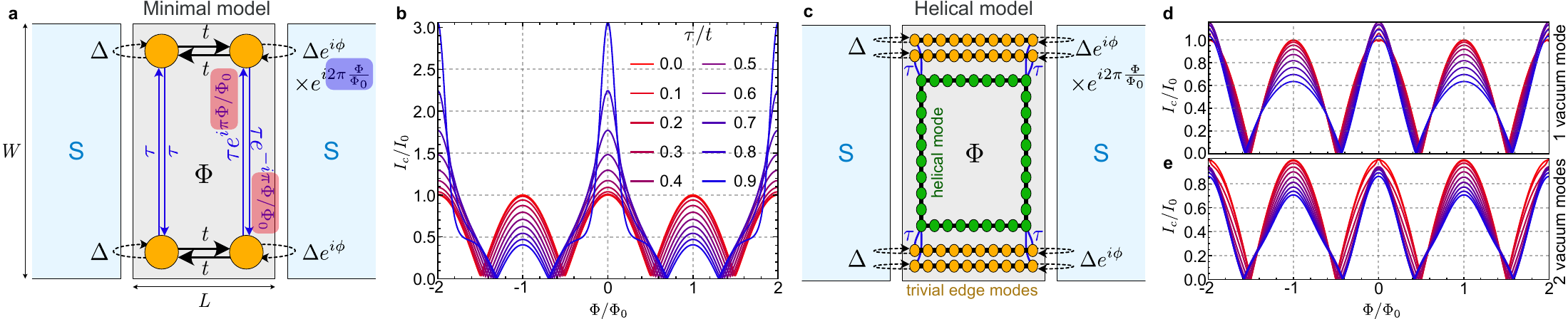}
  \end{center}
\caption{\textbf{Theory of the Fraunhofer even-odd effect from helical edge states.} \textbf{a} Sketch of the four-site minimal model, in terms of the induced superconducting pairings at the corners (with amplitude $\Delta$, phase $\phi$, magnetic flux $\Phi$ and flux quantum $\Phi_0$), intra-edge hoppings $t$ and inter-edge hoppings $\tau$, with their Peierls phases in the gauge $A_y=Bx$. The origin of coordinates is chosen at the bottom-left corner. \textbf{b} Critical current $I_c(\Phi)$, normalized to a fixed $I_0$, of the minimal model exhibiting an even-odd modulation for $\tau\neq 0$. \textbf{c} Sketch of a more elaborate model, including in each spin sector one or more trivial edge modes (yellow) and one helical mode (green) flowing around a gapped bulk. \textbf{d,e} Corresponding $I_c(\Phi)$ for one and two trivial edge modes, respectively, exhibiting the even-odd effect when trivial and helical modes become coupled by a hopping $\tau$ at the corners.}
\label{fig4}
\end{figure*}

\begin{figure*}
  \begin{center}
    \includegraphics[width=\textwidth]{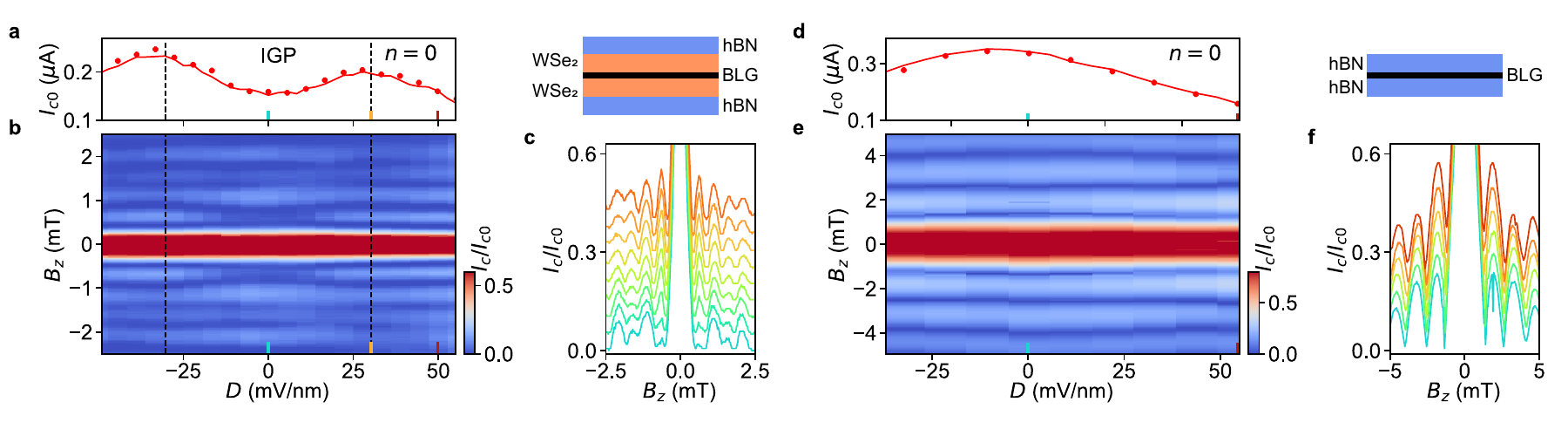}
  \end{center}
\caption{\textbf{Even-odd effect in superconducting quantum interferometry.} \textbf{a} Zero-field critical current $I_{c0}$ at zero carrier density for Dev~B extracted from Fig.~\ref{fig3}b (solid line) and from Fig.~\ref{fig5}b (circles) as a function of displacement field $D$. \textbf{b} Normalized critical current $I_c/I_{c0}$ as a function of perpendicular magnetic field $B_z$ and $D$. An even-odd effect in the SQI oscillatory pattern can be observed only within the IGP (marked by two vertical dotted lines). \textbf{c} $I_c/I_{c0}(B_z)$ at different $D$ ranging from 0 to 50~mV/nm, with a vertical shift of 0.05 per curve. 
\textbf{d-f} Same as Fig. \ref{fig5}a-c but for Dev~C where a BLG is encapsulated in hBN without WSe$_2$. In \textbf{d} the solid line is extracted from the supercurrent measurements in the Appendix D.
}
\label{fig5}
\end{figure*}

Given its importance for the interpretation of our SQI results, we now provide a theoretical analysis of how $2\Phi_0$ harmonics arise in $I_c(\Phi)$ as a result of electrons and holes circulating around the sample. We compute $I_c$ in a simple model containing the minimal ingredients to capture the effect (see the Appendix G for further details). The JJ with a gapped central region is abstracted into just four sites; one per corner of the BLG region. The parent superconductor induces on the $j$-th corner a pairing potential $\Delta e^{i\phi_j}$, see Fig.~\ref{fig4}a. The pairing phases $\phi_j$ depend on the magnetic flux $\Phi$ through the BLG region, which is described by a gauge field $A_y=B_z x$. This makes the tight-binding Hamiltonian $H$ of the four sites strictly $\Phi_0$-periodic in $\Phi$. Next, we add two different hopping amplitudes between the corners, representing the existence of edge channels flowing around the central region. These hoppings acquire a Peierls phase induced by $A_y$. The horizontal hoppings along the vacuum edges are denoted by $t$. They have zero Peierls phase, and thus preserve the $\Phi_0$ periodicity of the Hamiltonian. The vertical hopping along the left NS interface is denoted by $\tau$, again with zero Peierls phase since it is located at $x=0$. On the right NS interface at $x=L$ the hopping has the same modulus $\tau$, but its Peierls phase is now $e^{\pm i \pi \Phi/\Phi_0}$, depending on the direction. This term makes the Hamiltonian $2\Phi_0$-periodic when $\tau\neq 0$.

The critical current $I_c$ may be computed by maximizing the Josephson current $I=(2e/\hbar)\partial_\phi F$ versus $\phi$, where $F$ is the free energy computed from $H$. The resulting $I_c$, shown in Fig.~\ref{fig4}b, therefore inherits the periodicity of $H$ with $\Phi$. For a small $\Delta/t$ and a small but finite $\tau/t$, the resulting critical current is accurately approximated by~\cite{Baxevanis_even_2015} $I_c\sim|\cos(\pi\Phi/\Phi_0)+f|$ for a positive $0\leq f<1$ that grows with $\tau$, see the Appendices F and H. A finite $f$ results in an even-odd modulation of the $\tau=0$ SQUID-like pattern. As $\tau$ approaches $t$, $I_c$ develops higher harmonics that deviate from this simple expression. The transport processes enabled by the coupling $\tau$ along the NS interfaces, i.e., by Josephson currents looping around the normal region, lead to the even-odd effect in the Fraunhofer pattern.

We now improve the minimal model by adopting a more microscopic description with several trivial transport channels flowing along vacuum edges~\cite{allen_2016_spatially,zhu_edge_2017}, and one helical mode flowing all around the BLG junction, as expected from the band-inverted phase~\cite{zaletel_gate-tunable_2019,island_spin-orbit_2019,san-jose_majorana_2022}. A coupling $\tau$ between vacuum and helical states enables an inter-edge scattering mechanism, see Fig.~\ref{fig4}c. This time the Peierls phase is incorporated into the tight-binding discretization of the different modes. Like in the minimal model, the inter-mode coupling and the superconducting proximity effect are both assumed to take place at the corners of the normal sample, whose bulk is again assumed to be completely gapped. Although these simplifying assumptions are not necessarily satisfied by the experimental samples, they are enough to confirm that the conclusions drawn from the minimal model still hold in a more generic situation, with propagating helical modes in place of a direct inter-edge hopping and with multiple trivial vacuum edge modes. The results for $I_c$ in the case of one and two vacuum edge modes are shown in Fig.~\ref{fig4}(d,e). We once more find a $\Phi_0$-periodic SQUID-like pattern at $\tau=0$ (in red), representative of the non-inverted phase. By switching on the coupling $\tau$, an even-odd modulation arises. This holds true also for higher number of vacuum modes. The main difference with respect to the minimal model is the behavior when $\tau\to t$, which is now less drastic.

Keeping these theoretical results in mind, we measure the SQI patterns as a function of $D$ at $n=0$ (Fig.~\ref{fig5}b,c). As predicted by our model, we observe an even-odd effect in the $I_c (B_z)$ oscillations inside the IGP. Unlike in standard Fraunhofer and SQUID patterns, odd lobes are smaller than the adjacent even lobes, see for instance the lower curve of Fig.~\ref{fig5}c. Such modulation is lost as we enter the BIP (upper curves), wherein an SQUID-like pattern is gradually recovered as $|D|$ is increased, probably originating from trivial and decoupled vacuum edge modes (see Appendix E). To confirm that the even-odd effect is associated to superconducting helical modes and not to bulk supercurrents, we perform control experiments on a bare hBN/BLG/hBN JJ (Dev~C), where we expect no induced SOC and hence no helical modes. Indeed, we do not observe any even-odd effect in the control SQI patterns although the trivial edges are still present at higher $D$ (Fig.~\ref{fig5}e,f and Appendix E).

Even-odd modulated SQI patterns have been experimentally reported in various 2D systems \cite{pribiag_edge_2015,bocquillon_gapless_2017,de_vries_he_2018,de_vries_crossed_2019}. In InAs- and InSb-based planar JJs~\cite{de_vries_he_2018,de_vries_crossed_2019}, it was proposed that the effect stems from crossed Andreev reflection connecting the trivial edges of the JJs via conducting NS interfaces. Unfortunately, its origin was not fully clarified in the InAs and InSb experiments. In contrast, our symmetrically encapsulated BLG junctions offer a very natural candidate for an efficient inter-edge coupling mechanism, in the form of helical modes flowing along the two NS boundaries in the IGP. The even-odd effect then acquires a special significance in our experiment, as a direct probe of the weak-topological nature of the inverted gap, and as a demonstration of how $D$ can control the emergence of helical modes.

Our work provides the first experimental evidence of  supercurrent flow along helical edges in graphene. The clear $2\Phi_0$-periodic signature in our SQI experiment suggests an enticing prospect of detecting topological $4\pi$-periodic current-phase relations in this system~\cite{Beenakker_fermion_2013}, which requires the investigation of the a.c. Josephson effect \cite{bocquillon_gapless_2017,wiedenmann_4pi_2016,Laroche_2019_observation} and/or non-equilibrium effects at finite bias in a.c. supercurrent \cite{Haidekker_2020_superconducting}. The observation of gate tunable helical-edge-mediated supercurrents opens a promising new avenue \cite{san-jose_majorana_2022} towards topological superconductivity and Majorana physics in van der Waals materials.
\\

\textbf{Data availability}
Raw data, analysis scripts and simulation code for all presented figures
are available at Refs. \onlinecite{prasanna_rout_2023_7944462,fernandopenaranda_2023_7941266}.

\textbf{Author contributions}
P. R. and N. P. fabricated and measured the devices, and analysed the experimental data. S. G. conceived and coordinated the experiment. F. P., E. P. and P. S-J. developed the theoretical interpretation and models, and F. P. performed the simulations. All authors contributed to writing the manuscript.

\textbf{Acknowledgements}

We thank Joshua Island for discussions regarding heterostructure assembly and Tom Dvir for comments on the manuscript. This research was supported by an NWO Flagera grant and TKI grant of the Dutch Topsectoren Program, the Spanish Ministry of Economy and Competitiveness through Grants PCI2018-093026 (FlagERA Topograph) and PID2021-122769NB-I00 (AEI/FEDER, EU), and the Comunidad de Madrid through Grant S2018/NMT-4511 (NMAT2D-CM).

\appendix

\renewcommand{\thefigure}{S\arabic{figure}}
\setcounter{figure}{0}

\section{Device fabrication}
We exfoliate flakes of bilayer graphene (BLG), graphite (5-15 nm), tungsten diselenide WSe$_2$ (10-35 nm) and hexagonal boron nitride hBN (20-55 nm) on different SiO$_2$/Si substrates using Scotch tape. Then the stacks of hBN/WSe$_2$/BLG/WSe$_2$/hBN/graphite are assembled using the van der Waals dry-transfer using polycarbonate (PC) films on Polydimethylsiloxane (PDMS) hemispheres. The flakes are picked up at 110$^\circ$ C and the PC is melted at 180$^\circ$ C on SiO$_2$/Si substrates in the end. After dissolving the melted PC in N-Methyl-2-pyrrolidone (NMP), the stacks are checked with atomic force microscopy (AFM). 

\begin{figure*}
  \begin{center}
    \includegraphics[width=0.6\textwidth]{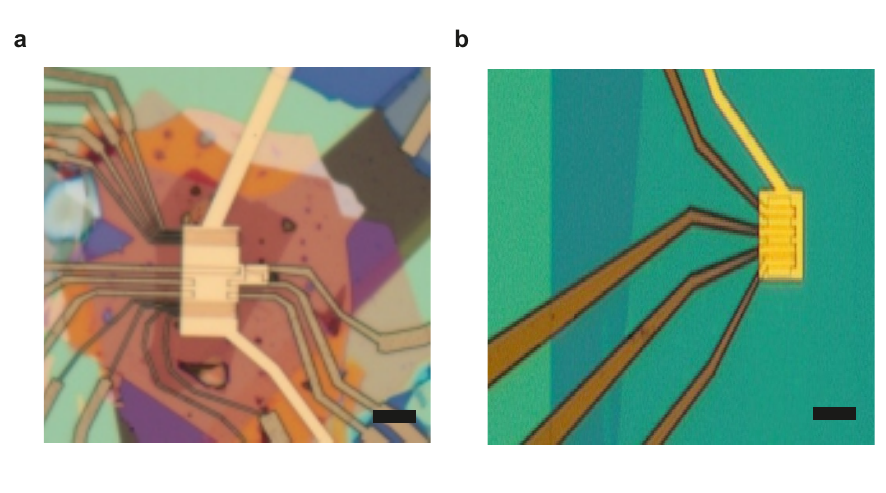}
  \end{center}
\caption{Optical image of JJs fabricated on (a) hBN/WSe$_2$/BLG/WSe$_2$/hBN/graphite and (b) hBN/BLG/hBN/graphite stacks. Scale bar is 5 $\mu$m.}
\label{Fig:image}
\end{figure*}

To fabricate the Josephson junctions (JJs) we spin coat a bilayer of 495 A4 and 950 A3 PMMA resists at 4000 rpm and bake at 175$^\circ$ C for 5 minutes, after each spinning. After the $e$-beam lithography patterning, the resist development is carried out in a cold $\mathrm{H_2O}$:IPA 3:1 mixture. We perform a reactive ion etching step using CHF$_3$/$\mathrm{O_2}$ mixture (40/4 sccm) at 80 $\mu$bar with a power of 60 W to etch precisely through the top hBN and WSe$_2$. Then we deposit NbTi (5nm)/NbTiN (110 nm) by dc sputtering for superconducting edge contacts and lift-off in NMP at 80$^\circ$ C. In order to isolate the top gate metal from the ohmic, we deposit a 30~nm dielectric film of Al$_2$O$_3$ using atomic layer deposition (ALD) at 105$^\circ$ C. Finally we define the top gate by e-beam lithography, deposit Ti (5~nm)/Au (120~nm), followed by lift-off (Fig.~\ref{Fig:image}).

In this work we measure three devices: two JJs with the hBN/WSe$_2$/BLG/WSe$_2$/hBN stack (Dev A, Dev B) and one JJ with hBN/BLG/hBN (Dev C). The two former JJs are 7 $\mu$m wide and their respective lengths are 3.7~$\mu$m (Dev A) and 300~nm (Dev B). Dev C is 3~$\mu$m wide and 300~nm long. 

\section{Measurements}
All measurements are performed in a dilution
refrigerator with a base temperature of 40 mK. Four-probe transport measurements are performed with a combination of DC and AC current bias scheme. To determine critical current we have employed a voltage switching detection method with current source and critical current measurement modules developed at TU Delft.\cite{ivvi} The magnetic fields are applied by a 3D vector magnet, which enables us to align the field within $\pm$5$^\circ$ accuracy.

\section{$\text{WSe}_2$/BLG/$\text{WSe}_2$ devices}
\begin{figure}
  \begin{center}
    \includegraphics[width=\columnwidth]{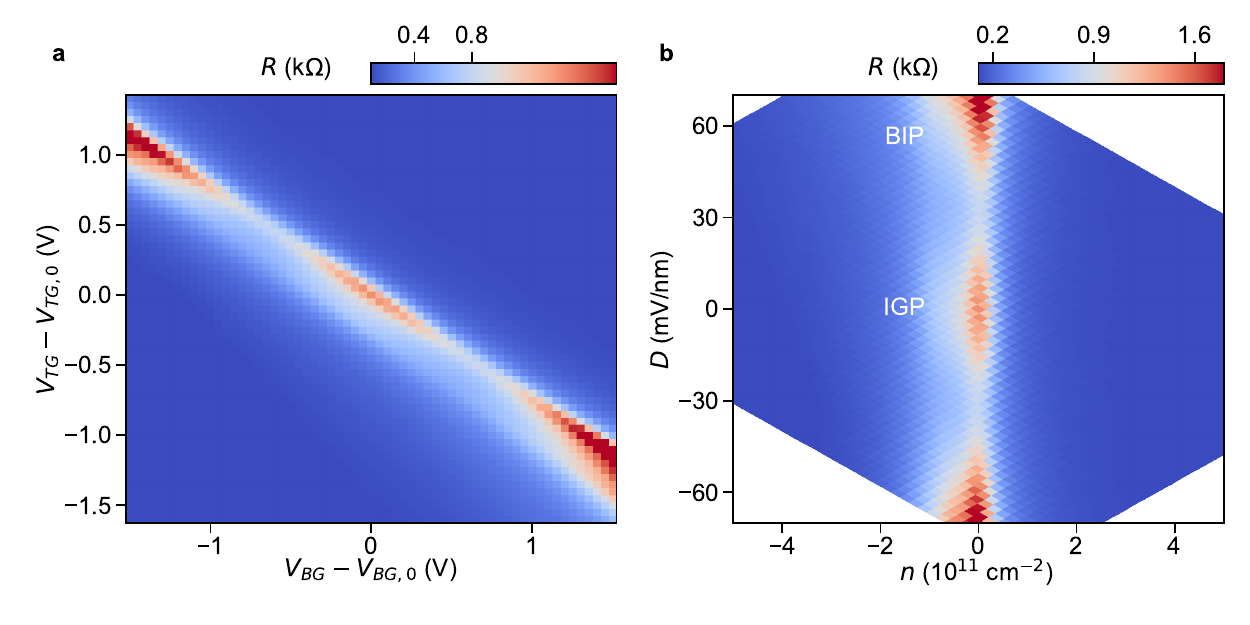}
  \end{center}
\caption{(a) Normal-state resistance $R$ measured as a function of $V_{BG}$ and $V_{TG}$ at 3.3~K for Dev A. Here $V_{BG,0} = 0.5$~V and $V_{TG,0} = -1.4$~V.  (b) $R$ replotted as a function of $n$ and $D$ using the capacitances $C_{BG} = 4.42\times 10^{-4}$~F and $C_{TG} = 5.765\times 10^{-4}$~F.}
\label{Fig:DevA}
\end{figure}


\begin{figure}
  \begin{center}
    \includegraphics[width=\columnwidth]{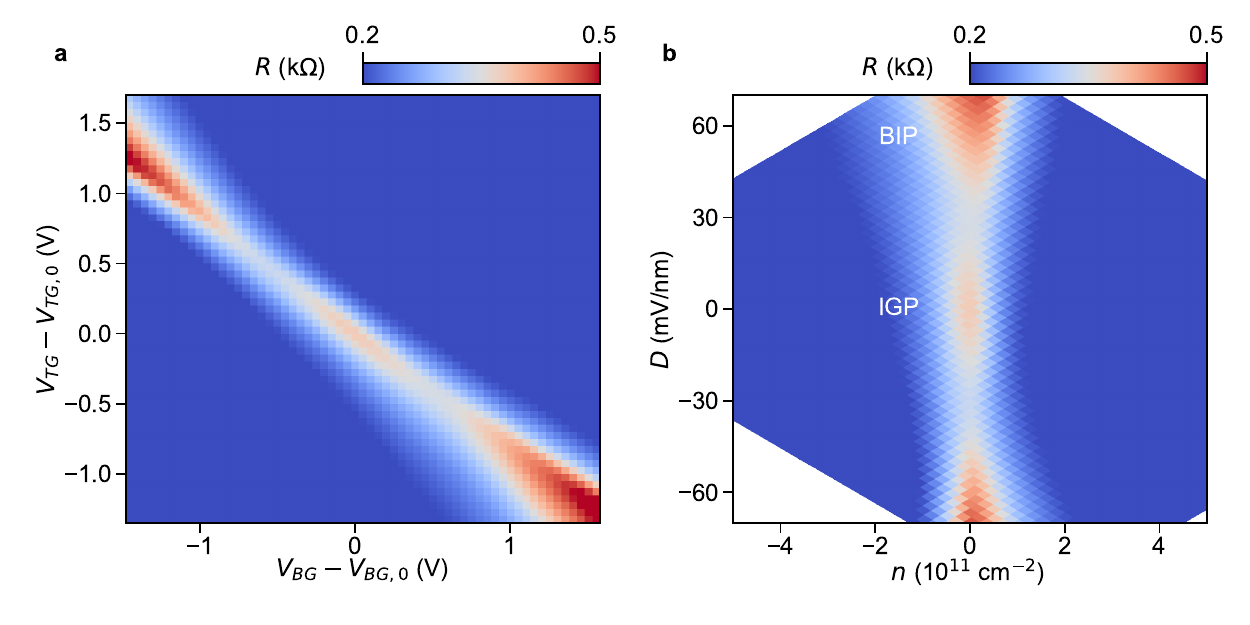}
  \end{center}
\caption{(a) $R$ measured as a function of $V_{BG}$ and $V_{TG}$ at 3.3~K for Dev B. Here $V_{BG,0} = 0.5$~V and $V_{TG,0} = -1.675$~V. (b) $R$ replotted as a function of $n$ and $D$ using the capacitances $C_{BG} = 4.42\times 10^{-4}$~F and $C_{TG} = 5.765\times 10^{-4}$~F.}
\label{Fig:DevB}
\end{figure}

Figure~\ref{Fig:DevA}a shows the dual gate (back gate $V_{BG}$ and top gate $V_{TG}$) map of the normal-state resistance $R$ for Dev A. In Fig.~\ref{Fig:DevA}b, $R$ is replotted as a function of $n$ and $D$. This clearly shows a resistance maximum near $n = 0 = D$ indicating the inverted-gap phase (IGP). Similar resistance maps are measured for Dev B (Fig.~\ref{Fig:DevB}).

\begin{figure}
  \begin{center}
    \includegraphics[width=\columnwidth]{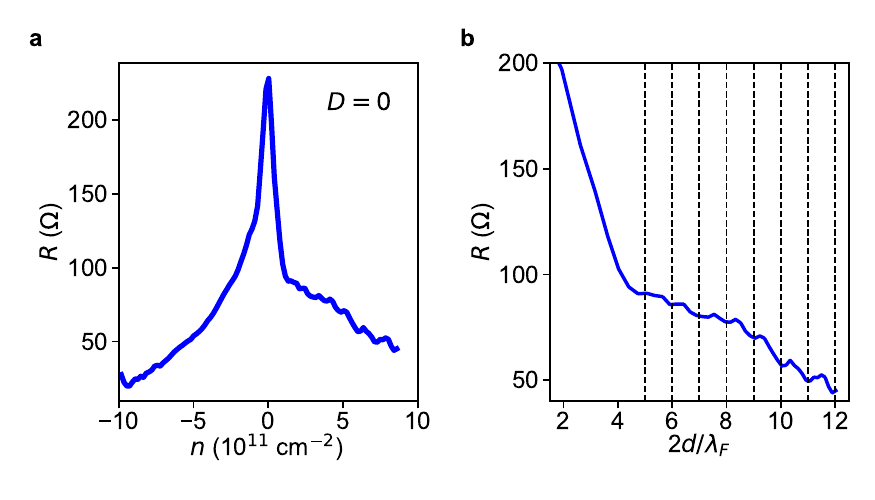}
  \end{center}
\caption{(a) $R$ measured as a function of carrier density $n$ at displacement field $D = 0$ for Dev C. (b) $R$ replotted as a function of $2L/\lambda_F$. The vertical dashed lines mark the period of the Fabry-Perot oscillations.}
\label{Fig:FP}
\end{figure}

The existence of Fabry-Perot (FP) interferences in the normal-state resistance curve shows that Dev B is a ballistic JJ (Fig.~\ref{Fig:FP}). The FP resonances appear when the condition $2d = m\lambda_F$ is satisfied, where $d$ is the cavity length, $m$ is an integer and $\lambda_F$ is the Fermi wavelength. Figure~\ref{Fig:FP}b shows the resistance oscillations with a period of $2d/\lambda_F$ for $d = 325$~nm.

\section{Bare BLG device}

\begin{figure*}
  \begin{center}
    \includegraphics[width=0.9\textwidth]{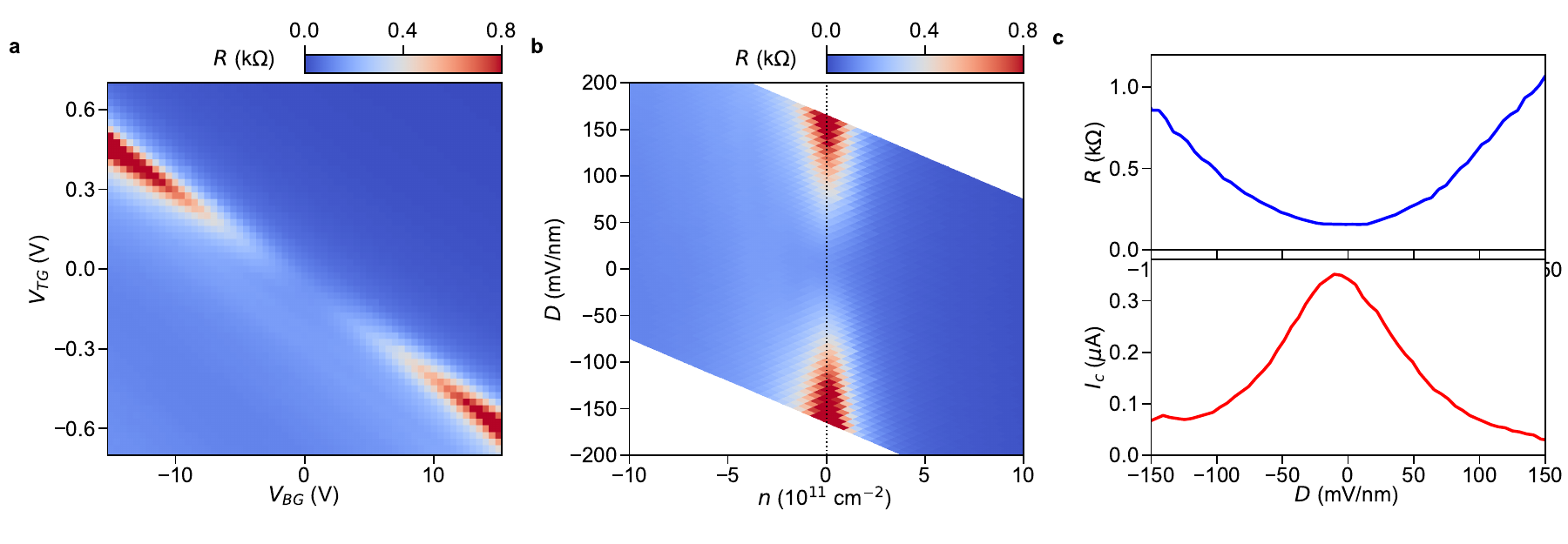}
  \end{center}
\caption{(a) $R$ measured as a function of $V_{BG}$ and $V_{TG}$ at 3.3~K for Dev C. Here $V_{BG,0} = 0$~V and $V_{TG,0} = -0.07$~V.(b) $R$ replotted as a function of $n$ and $D$ using the capacitances $C_{BG} = 0.96\times 10^{-4}$~F and $C_{TG} = 2.75\times 10^{-3}$~F. (c) Resistance line cut $R(D)$ of (b) at $n = 0$ (top panel) and critical current $I_c(D)$ for $n = 0$ at 40~mK (bottom panel).}
\label{Fig:BLG2}
\end{figure*}

Figure~\ref{Fig:BLG2} shows the resistance $R$ as a function of $V_{BG}$ and $V_{TG}$ (a) as well as a function of $n$ and $D$ (b) of the BLG JJ without any WSe${}_2$ encapsulation. Clearly we observe an increase in $R$ with increasing $|D|$ due to the opening of a trivial gap (Fig. \ref{Fig:BLG2}c). In addition, the critical current $I_c$ gradually reduces with increasing $|D|$ as expected from the $R(D)$ behaviour.

\section{SQI pattern  at high doping}
\begin{figure}
  \begin{center}
    \includegraphics[width=0.8\columnwidth]{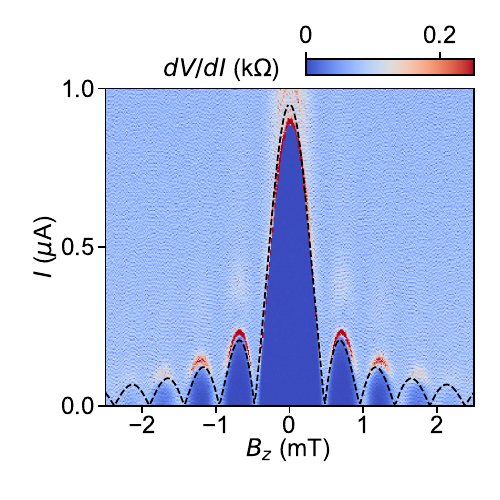}
  \end{center}
\caption{Superconducting quantum interference pattern $I_c$($B_z$) for Dev~B at $n = 3\times10^{11}$ cm$^{-2}$ and $D =0$, that follows the standard Fraunhofer curve (dashed line).}
\label{Fig:SQI_highdoping}
\end{figure}

Figure~\ref{Fig:SQI_highdoping} displays a Fraunhofer superconducting quantum interferometry (SQI) pattern for Dev~B at $n = 3\times10^{11}$~cm$^{-2}$ and $D =0$. The SQI for a JJ with homogeneous supercurrent flow is of the form: $I_c(B_z) = I_{c,0} \sin(\pi\Phi/\Phi_0)/ (\pi\Phi/\Phi_0)$, where $I_{c,0}$ is the zero-field critical current and $\Phi_0=h/2e$ is the (superconducting) flux quantum. The magnetic flux $\Phi$ passing through the JJ is given by $B_z/LW$, where $L$ and $W$ are the length and width of the junction, respectively. However, the flux focusing increases the effective length. For our JJs the magnetic penetration depth of NbTiN ($\approx$ 400~nm)\cite{Kroll_2019_magnetic} is larger than the film thickness ($\approx$110~nm) and the width of the electrode ($l \approx$300~nm). Thus, the effective JJ length is $L +l =$ 600 nm. Using this value we plot the standard Fraunhofer curve in~Fig.~\ref{Fig:SQI_highdoping}, which matches quite well the measured SQI pattern.

\section{SQI patterns for $n = 0$ at larger $D$}

\begin{figure*}
  \begin{center}
    \includegraphics[width=\textwidth]{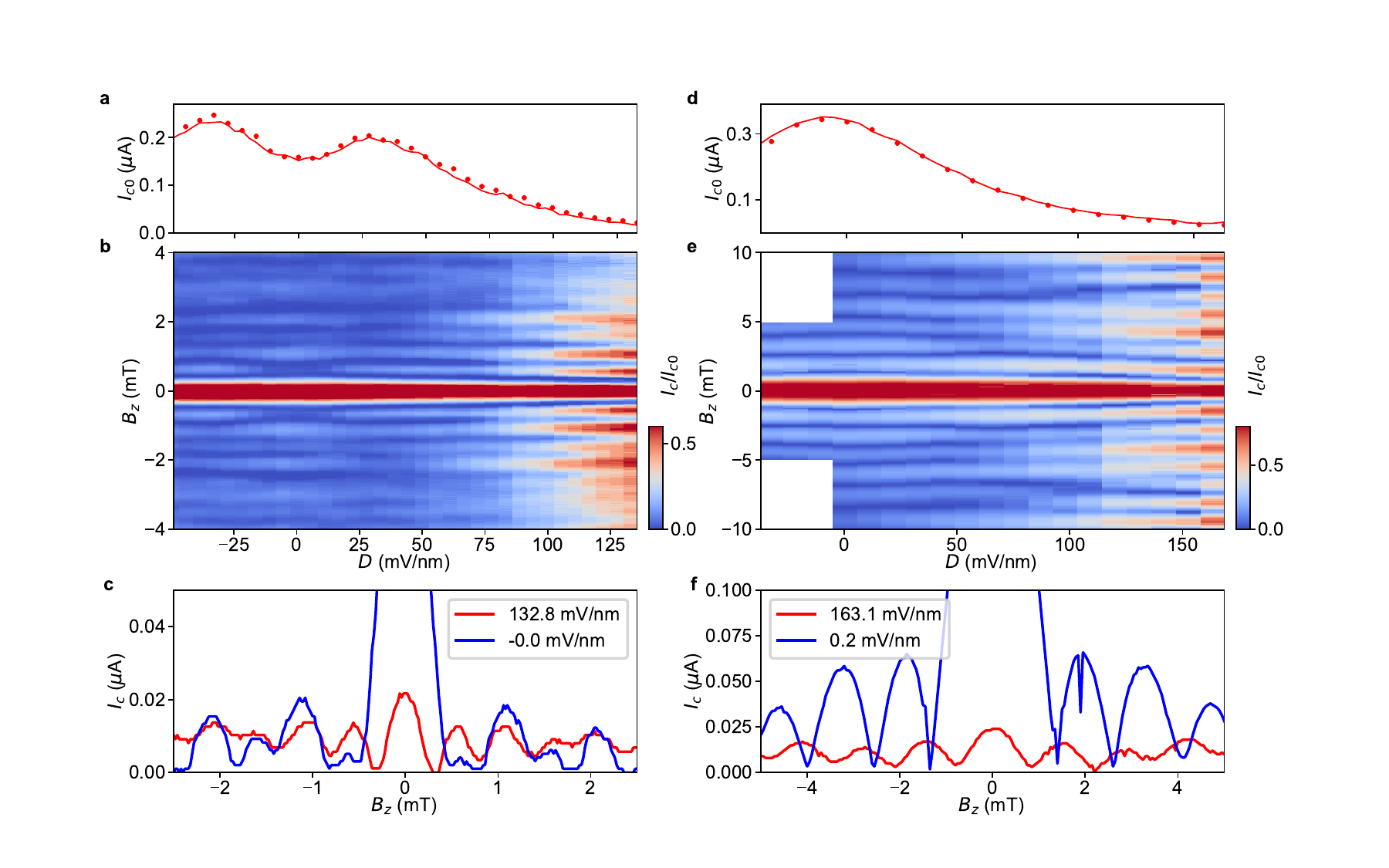}
  \end{center}
\caption{(a) Zero-field critical current $I_{c0}$ extracted from Fig.~\ref{fig3}b (solid line) and a line cut at $B_z = 0$ of (b) (solid circles) as a function of $D$ for Dev~B. (b) SQI pattern as a function of $B_z$ and $D$ at $n = 0$ for Dev B. (c) Line cuts of (b) at $D = $~0 mV/mm and $D =$~132.8~mV/mm. (d) Zero-field critical current $I_{c0}$ extracted from Fig.~\ref{Fig:BLG2}c (solid line) and a line cut at $B_z = 0$ of (e) (solid circles) as a function of $D$ for Dev~C. (e) SQI pattern as a function of $B_z$ and $D$ at $n = 0$ for Dev C. (f) Line cuts of (e) at $D =$~0.2 mV/mm and $D =$~163.1 mV/mm.}
\label{Fig:FH-SQI}
\end{figure*}

Figure~\ref{Fig:FH-SQI} presents the SQI patterns measured at larger $D$ values and $B_z$ for Dev~B and Dev~C. As discussed in main text, the even-odd effect is only seen in the WSe$_2$-encapsulated JJ. For large $D$ values we observe symmetric SQUID-like SQI patterns for both JJs indicating the presence of trivial edge states. These edge states have been previously observed in BLG JJs~\cite{allen_2016_spatially,zhu_edge_2017}. 


\section{Formula for the critical current}
Neglecting phase fluctuations, the critical current across a two-terminal JJ can be written as
\beq I_c(\Phi) = \max_{2\pi \geq \phi \geq 0} \big| I(\phi, \Phi) \big|, \label{Icmax} \eeq
where $I(\phi, \Phi)$ is the supercurrent at a given superconducting phase difference, $\phi$, and magnetic flux, $\Phi$, threading the junction area. In the zero-temperature limit, $I(\phi, \Phi)$ can in turn be expressed purely as a function of spectral quantities, namely the system's free energy $F$, as
\beq I(\phi, \Phi) =  \frac{2e}{\hbar} \frac {d }{d\phi}F(\phi, \Phi) = \frac{2e}{\hbar} \frac {d}{d\phi} \sum_{n\in \text{occ}} E_n(\phi, \Phi), \label{Ic}\eeq
written as the sum of the energies $E_n$ over all occupied eigenstates of the junction Hamiltonian $H$. These correspond to negative eigenvalues of the Bogoliubov-de Gennes (BdG) Hamiltonian.

\section{Construction of the four-site minimal model}
\begin{figure*}
  \begin{center}
\includegraphics[width=0.7\textwidth]{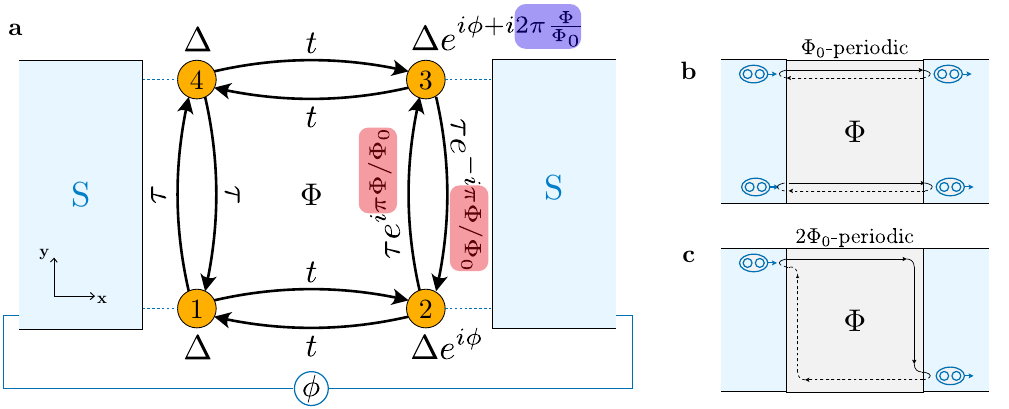}
\end{center}
\caption{ \textbf{(a)} Sketch of a minimal model for the Josephson junction (JJ) with vacum edge states, defined by hoppings $t$, and an inter-edge coupling, defined by hoppings $\tau$. Sites 1, 4 and 2, 3 are coupled to the left and right superconducting leads, respectively, which have a phase difference $\phi$. In addition, a finite magnetic flux $\Phi$ induces Peierls and superconducting phases at the hopping amplitudes and onsite pairing terms, respectively. Crucially, the Peierls phase on those hoppings connecting different edges (in red) halves the flux-induced superconducting phase at site 3 (in blue). The induced superconducting pairing amplitudes at the corners are $\Delta$. \textbf{(b)} Supercurrent trajectories that contribute to a $\Phi_0$-periodic SQUID-like Fraunhofer pattern. \textbf{(c)} Trajectories encircling the normal region enabled by inter-edge coupling, that contribute with $2\Phi_0$-periodic critical currents.}
\label{fig:foursite}
\end{figure*}

In order to get physical insight into the processes that may give rise to an even-odd modulation in $I_c(\Phi)$, we propose, prior to more elaborate calculations, the following minimal tight-binding model for the junction. It consists of only four sites located at the corners of the junction, see Fig. \ref{fig:foursite}a. These four sites represent the normal region between the superconductors, and are threaded by a magnetic flux $\Phi$. Within this model, the sites labelled as $1$ and $4$, and $2$ and $3$ are contacted to the left and right superconducting leads, respectively, in a JJ geometry. The resulting BdG Hamiltonian for a single spin species reads
\beq H = \begin{pmatrix} H_0 & H_\Delta^+\\ H_\Delta & -H_0^* \end{pmatrix}, \label{bdgH}
\eeq
written in the Nambu basis $(\boldsymbol{c}_{\sigma},\boldsymbol{c}^\dagger_{\bar \sigma})^T$ where  $\boldsymbol{c}_{\sigma}= (c_{1 \sigma},c_{2 \sigma},c_{3 \sigma},c_{4 \sigma})$ and
\beqa
H_0 &=& \begin{pmatrix} 0 & t & 0 & \tau\\ t & 0 & \tau e^{-i\pi \Phi/\Phi_0} & 0\\ 0 & \tau e^{i\pi \Phi/\Phi_0} & 0 & t\\ \tau & 0 & t &0 \end{pmatrix} \ \ \text{,} \\
H_\Delta &=& \Delta \begin{pmatrix} 1 & 0 & 0 & 0\\ 0 & e^{i \phi} & 0 & 0\\ 0 & 0 & e^{i \phi+i 2\pi \Phi/\Phi_0} & 0\\ 0 & 0 & 0 &1 \end{pmatrix}.
\eeqa
$H_0$ and $H_\Delta$ are $4\times4$ matrices corresponding to the normal Hamiltonian of the particle sector and the onsite superconducting electron-hole pairing terms induced by the leads, respectively. $\Delta$ and $\phi$ are the induced pairing amplitude and phase from the parent superconductors, respectively. $t$ and $\tau$ are the hopping amplitudes between sites along the vacuum edge interfaces ($1\leftrightarrow 2$ and $3\leftrightarrow4$) and the NS interfaces ($1\leftrightarrow 4$ and $2\leftrightarrow3$), respectively. Thus, the ratio $\tau/t$ is a key quantity that controls the efficiency of the inter-edge coupling (between the top and bottom edges).
 
 For the sake of simplicity, we assume that the system is spin degenerate and, thus, Eq. \eqref{bdgH} can be regarded as half the BdG Hamiltonian of a spinful model. The previous assumption is justified as long as the magnetic length at $\Phi = \Phi_0$ is the largest spatial scale of the problem, which seems a good approximation for large-area devices. 
 
The vector potential is chosen in the gauge $\boldsymbol A = A_y \boldsymbol y = \frac{\Phi}{L W} x \boldsymbol y$, where $L$ and $W$ are the length and width of the junction, respectively. The origin of coordinates is chosen at site 1. The magnetic flux introduces a position-dependent modulation of the pairing phase, following $\Delta_{\mathrm{parent}}(\mathbf{r}) = \Delta_{\mathrm{parent}}(\mathbf{r}_0)\exp(2i\int_{\mathbf{r}_0}^\mathbf{r} \mathbf{A}\cdot\mathbf{dr})$, where $\mathbf{r}_0$ is any given point inside the superconductor and the integral path is taken inside it. Taking $\mathbf{r}_0$ in the left lead, adjacent to site 1, this introduces a phase $\phi$ at site 2 (junction phase difference picked along the outer loop between left and right) and $\phi+2 \pi \Phi$ at site $3$. The latter makes the Hamiltonian $\Phi_0$-periodic in $\Phi$. In addition, $\Phi$ enters $H_0$ as a Peierls phase in the hoppings. In the chosen gauge, the Peierls phase is non-zero only for hoppings connecting sites $2\leftrightarrow 3$. The Peierls phase makes $H$ $2\Phi_0$-periodic in $\Phi$, in contrast to the $\Phi_0$-periodicity from the pairing phases. From a different point of view, electrons and holes will pick up an Aharonov phase $\pm \pi \Phi/\Phi_0$ when circulating around the normal region (possible only if $\tau\neq 0$, which enables inter-edge scattering), which will produce a beating when added to the $2\pi\Phi/\Phi_0$ phase of a standard SQUID discussed in the main text.
Thus, and despite its extreme simplicity, the four-site model allows us to identify inter-edge scattering as the key mechanism behind the even-odd effect.

\begin{figure}
  \begin{center}
\includegraphics[width=\columnwidth]{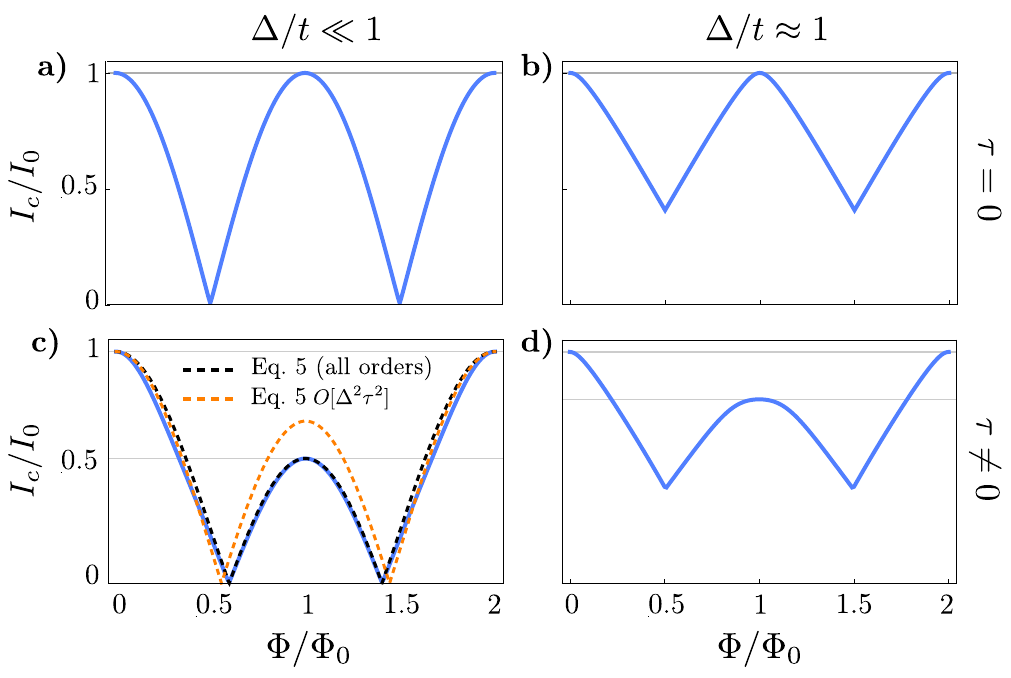}
\end{center}
\caption{Regimes of the minimal four-site model. $\Delta/t = 0.1$ (a, c), $\Delta/t = 0.9$ (b, d), $\tau/t = 0$ (a, b), $\tau/t = 0.5$ (c, d).}
\label{fig:foursitelimits}
\end{figure}

\section{Regimes of the four-site model}

Here we compute the critical current $I_c(\Phi)$ using Eqs. \eqref{Icmax} and \eqref{Ic} and the spectrum of Eq. \eqref{bdgH} in four different regimes. The results are shown in Fig. \ref{fig:foursitelimits}. The left and right columns correspond to the $\Delta/t \ll 1$ and $\Delta/t \approx 1$ regimes, respectively, whereas the top and bottom row contains the cases of forbidden or allowed inter-edge coupling controlled by $\tau/t$. 
In the absence of inter-edge coupling, Fig. \ref{fig:foursitelimits}(a,b), the spectrum has $\Phi_0$-periodicity whose oscillations with $\Phi$ may or may not touch zero at half-integer fluxes depending on the value of $\Delta/t$. A small $\Delta/t$ ratio yields a complete suppression of $I_c$ at half-integer normalized flux (a) as opposed to (b). 

The beating of the $\Phi_0$-periodic modulation comes from electron/hole trajectories encircling the sample [see Fig. \ref{fig4}(d, e)] and is enabled by a finite $\tau/t$. In Fig. \ref{fig:foursitelimits}c, the experimentally relevant situation, the exact $I_c(\Phi)$ functional form, in blue, can be approximated by \beq I_c(\Phi) \approx \frac{I_c(0)+I_c(\Phi_0)}{2I_c(0)} \left| \cos(\pi \frac{\Phi}{\Phi_0}) + \frac{I_c(0)-I_c(\Phi_0)}{I_c(0)+I_c(\Phi_0)}\right|, \label{approximation} \eeq depicted as a dashed black line, similar to Eq. (18) of Ref. \onlinecite{Baxevanis_even_2015} obtained in the framework of a network (transfer-matrix) model. This approximation becomes exact, however, only in the limit of small $\Delta/t$ and $\tau/t$. In the general case, in contrast, the critical current receives contributions from all possible processes without spin mixing that lead to the net transfer of Cooper pairs across the junction and, therefore, involve an arbitrary number of loops around the sample and Andreev electron/hole conversions. To gain a better understanding of those that contribute to the $\Phi_0$-periodicity breaking, we focus on the leading processes up to order $(\Delta/t)^2$ and $(\tau/t)^2$ corresponding to two Andreev reflections, one at each NS interface, and quasiparticle trajectories that encircles the device at most once, as those depicted in Fig. \ref{fig:foursite}(b, c). In this $\Delta/t,\tau/t \ll 1$ limit, the critical currents at $\Phi = 0$ and $\Phi = \Phi_0$ read:
\beq I_c(0) \approx \frac{2e}{\hbar} \frac{16\Delta^2}{t} \left(1+\frac{\tau^2}{t^2}\right) \label{Ic0}, \eeq
and
\beq I_c(\Phi_0) \approx \frac{2e}{\hbar} \frac{16\Delta^2}{t} \left(1-\frac{\tau^2}{2t^2}\right) \label{Icphi0}. \eeq
Substituting Eqs. \eqref{Ic0} and \eqref{Icphi0} into Eq. \eqref{approximation} yields the dashed orange line in Fig. \ref{fig:foursitelimits}c that tends to the exact solution as $\tau/t \rightarrow 0$. Despite its deviation from the calculations to all orders (blue and dashed black lines in Fig. \ref{fig:foursitelimits}) as we relax the $\tau/t \ll 1$ condition, we conclude that loops of order $(\Delta/t)^2$ and $(\tau/t)^2$ are the most relevant within the hierarchy of inter-edge processes that leads to $2\Phi_0$-periodicity.

\section{Contribution to $I_c(\Phi)$ of $\mathcal{O}[\Delta^2 \tau^2]$ processes}

In this section we provide details of the derivation of Eqs. \eqref{Ic0} and \eqref{Icphi0}. We start by rewritting the BdG Hamiltonian of Eq. \eqref{bdgH} into a normal and a superconducting part as:
\beq
\frac{H}{t} = \tilde H_N+ \tilde H_S= \begin{pmatrix} \frac{H_0}{t} & 0 \\ 0 & - \frac{H_0^*}{t} \end{pmatrix} + \frac{\Delta}{t} \begin{pmatrix} 0 & \frac{H_\Delta^+}{\Delta} \\ \frac{H_\Delta}{\Delta} & 0 \end{pmatrix}. \label{perturbH}
\eeq
In the regime of interest for the experiment, i.e., that of Fig. \ref{fig:foursitelimits}c with $\Delta/t \ll 1$, $\tilde H_S$ can be regarded as a small perturbation to $\tilde H_N$ and, therefore, a perturbative treatment of the problem is justified. Let $|\Psi^{(0)}_i\rangle$ be the $i$'th eigensolution of \beq \tilde H_N |\Psi^{(0)}_i\rangle = E^{(0)}_i |\Psi^{(0)}_i\rangle, \eeq with energy $E^{(0)}_i$. Since the unperturbed Hamiltonian does not couple the electron and hole sectors we can distinguish said eigenstates by their electron/hole character:
\beq |\Psi^{(0)}_{i,e}\rangle = |e\rangle \otimes |\psi_{i,e}^{(0)} \rangle, \eeq 
\beq |\Psi^{(0)}_{i,h}\rangle = |h\rangle \otimes |\psi_{i,h}^{(0)} \rangle, \eeq
where $|e\rangle$ and $|h\rangle$ are the eigenstates of the Nambu projector $\mathbb P_{e/h} = \tau_0 \pm \tau_z$, respectively, and $ |\psi_{i,e}^{(0)} \rangle$ ($ |\psi_{i,h}^{(0)} \rangle$) the eigenstates of $H_0$  ($-H_0^*$). They obey the so-called biorthonormality conditions, which in this basis reads:
\beq \langle \Psi^{(0)}_{i,e} |  \Psi^{(0)}_{j,e} \rangle = \delta_{i,j}, \ \  \langle \Psi^{(0)}_{i,h} |  \Psi^{(0)}_{j,h} \rangle = \delta_{i,j}, \ \text{and} \  \langle \Psi^{(0)}_{i,h} |  \Psi^{(0)}_{j,e} \rangle = 0, \label{biortho} \eeq
and  \beq \langle \psi_{i,e}^{(0)}  |\psi_{j,e}^{(0)} \rangle=\langle \psi_{i,h}^{(0)}  |\psi_{j,h}^{(0)} \rangle = \delta_{i,j}. \eeq

Under standard perturbation theory the eigenstates and eigenvalues of $H$ can be expressed as:
\beq E_i = E_i^{(0)} + E_i^{(1)} + E_i^{(2)} + \mathcal O[(\Delta/t)^3]\ \ \ \text{and}\eeq
 \beq |\Psi_i \rangle = |\Psi^{(0)}_i \rangle + |\Psi^{(1)}_i \rangle + |\Psi^{(2)}_i \rangle + \mathcal O[(\Delta/t)^3].\eeq
where $E_i^{(n)}$ and  $|\Psi^{(n)}_i \rangle$ corresponds to the $n$'th order contribution to the $i$'th eigenstate $| \Psi_i \rangle$ with corresponding energy $E_i$ of the perturbed problem, respectively.

Substituting the previous expansion into the Schr\"odinger equation for the full BdG Hamiltonian yields:
\beq \left(H^{(0)} -E_i^{(0)}\right) |\Psi^{(1)}_i \rangle = \left(E_i^{(1)}-H^{(1)} \right) |\Psi^{(0)}_i \rangle, \label{firstorder} \eeq  and
\beq \left(H^{(0)} -E_i^{(0)}\right) |\Psi^{(2)}_i \rangle = E_i^{(2)} |\Psi^{(0)}_i \rangle+\left(E_i^{(1)}-H^{(1)} \right) |\Psi^{(1)}_i \rangle. \label{secondorder}\eeq
The artificial built-in redundancy of the Nambu representation for $\tilde H_N$ demands to use degenerate perturbation theory, as it reveals in the form of a two-fold degeneracy of the unperturbed energy levels, eventually splitted by the perturbation $\tilde H_S$. Following Ref. \onlinecite{Ling_perturbative_2021} we apply the following ansatz for the $i$-th two-fold degenerate subspace of the unperturbed system:
\beq |\Psi^{(0)}_{i}\rangle = \alpha_i | \Psi^{(0)}_{i,e}\rangle + \beta_i | \Psi^{(0)}_{i,h} \rangle, \eeq where $\alpha$ and $\beta$ are constants chosen to be real for simplicity. In a similar fashion, higher order eigenstates with $n>0$ can be decomposed as:
\beq |\Psi_i^{(n)} \rangle = \sum_{j\neq i} \alpha_j^{(n)} |\Psi_{j,e}^{(0)} \rangle+ \beta_j^{(n)} |\Psi_{j,e}^{(0)} \rangle. \eeq
Note that the $i\neq j$ constraint in the sum together with Eq. \eqref{biortho} imposes that the first and subsequent perturbed eigenstates are orthogonal to $|\Psi_i^{(0)} \rangle$, being consistent with the normalisation choice: $\langle \Psi_i^{(0)}|  \Psi_i \rangle = 1$. 

The first and second energy corrections to $E_i^{(0)}$ can be readily found by substituting these ansatszes into Eq. \eqref{firstorder} and Eq. \eqref{secondorder}, respectively, and solving the system of equations resulting from left multiplying it by $\langle \Psi_{i,e/h}^{(0)}|$. After some algebraic manipulations (see Ref. \onlinecite{Ling_perturbative_2021} for instance) we get: $E_i^{(1)} = 0$, since the BdG Hamiltonian is hermitian. On the other hand, the second-order corrections are given by:
\beq E_{i,e}^{(2)} =  \sum_{j\neq i} \frac{\langle \psi_{i,e} |\Delta|\psi_{j,h} \rangle \langle \psi_{j,h} |\Delta^*|\psi_{i,e} \rangle}{E_j^{(0)} - E_i^{(0)}}, \eeq and similarly
\beq E_{i,h}^{(2)} =  \sum_{j\neq i} \frac{\langle \psi_{i,h} |\Delta^*|\psi_{j,e} \rangle \langle \psi_{j,e} |\Delta|\psi_{i,h} \rangle}{E_j^{(0)} - E_i^{(0)}}, \eeq
with $i,j \in \{1,2,3,4\}$. Note that the bras and kets are the eigenstates of the normal $4\times4$ Hamiltonian $H_0/t$ instead of those of the $8\times8$ BdG $\tilde H_N$ in Eq. \eqref{perturbH}, a circumstance that greatly reduces the algebraic complexity. 

Now that we know the second order corrections to $E_i^{(0)}$, we can compute the free energy summing over all $i$'s corresponding to occupied states, and by virtue of Eqs. \eqref{Ic} and \eqref{Icmax} compute $I_c(\Phi)$. Taking the $\Phi \rightarrow 0$ and $\Phi \rightarrow \Phi_0$ limits of the resulting expression yields Eqs. \eqref{Ic0} and \eqref{Icphi0}.

\section{Construction of the multi-mode tight-binding model}

Going beyond the four-site model, we consider the following tight-binding Hamiltonian
\beq H_{TB} = H_\text{regions} + H_{\text{coupling}},
\label{HTB} 
\eeq 
where $H_{\text{regions}}$ comprises the Hamiltonians of three different regions [see sketch in Fig. \ref{fig4}c] and $H_{\text{coupling}}$ the couplings between them. The former reads \beq H_{\text{regions}}= H_{\text{IGP}} + H_{\text{vac}} + H_{\text{SC}}, \eeq
where $H_{\text{IGP}}$, $H_{\text{vac}}$, and $H_{\text{SC}}$ correspond, respectively, to the Hamiltonians of the IGP of BLG, two edge regions placed at the top and bottom interfaces to vacuum accounting for the expected formation of trivial edge channels, and the two s-wave superconducting leads at $x=0$ and $x=L$ with phase difference $\phi$ treated in mean-field. Setting the origin at the left bottom site of the NS interface, and using the same gauge for the vector potential as in the four-site model, the previous terms can be written as:
\beqa
H_\text{IGP} &=& \sum_{i,\sigma} \left( 2t - \mu_N \right) c_{i,\sigma}^\dagger c_{i,\sigma} \\
&&+ \sum_{\langle i,j \rangle,\sigma} \left( t e^{-i \pi  \frac{\Phi}{\Phi_0} \frac{x_i(y_j-y_i)}{LW}} c^\dagger_{i\sigma} c_{j\sigma} + \text{h.c.}\right),\nonumber
\eeqa
\beq H_{\text{vac}} = \sum_{i,\sigma} \left( 2t - \mu_{\text{vac}} \right) c_{i,\sigma}^\dagger c_{i,\sigma} + \sum_{\langle i,j \rangle,\sigma} \left( t c^\dagger_{i\sigma} c_{j\sigma} + \text{h.c.}\right), \eeq 
\beq H_{\text{SC}} =H^{L}_{\text{SC}} + H^{R}_{\text{SC}}, \eeq
\beq H^{L}_{\text{SC}} =\sum_{i\in L,\sigma} \left( 2t - \mu_{\text{SC}} \right) c_{i,\sigma}^\dagger c_{i,\sigma} - \sum_{i\in L} \left(\Delta c^\dagger_{i\uparrow} c^\dagger_{i \downarrow} +\text{h.c}\right),\eeq
\beqa
H^{R}_{\text{SC}} &=&\sum_{i \in R,\sigma} \left( 2t - \mu_{\text{SC}} \right) c_{i,\sigma}^\dagger c_{i,\sigma} \\
&&- \sum_{i \in R} \left(\Delta e^{i \phi} e^{-i 2\pi  \frac{\Phi}{\Phi_0}\frac{ x_i y_i}{LW}}  c^\dagger_{i\uparrow} c^\dagger_{i \downarrow} +\text{h.c}\right),\nonumber
\eeqa
where $c^\dagger_{i\sigma}$ creates at electron with spin $\sigma$ at site $i$, $\mu_{\text{N}}$, $\mu_{\text{vac}}$, and $\mu_{\text{SC}}$ are the chemical potentials at the helical, vacuum-edge, and superconducting regions, respectively, and $t=\frac{\hbar^2}{2 m a_0^2}$ where $a_0$ is the lattice constant and $m$ the electron mass. Note that in this Section $\Delta$ denotes the parent superconducting pairing amplitude (instead of the induced one used in the four-site model and the main text). Channels in the vacuum-edge region are assumed to be decoupled from each other and ballistic.
Finally, the helical region corresponding to the IGP only ``sees'' the superconductors through its coupling to the vacuum-edge region [see Fig. \ref{fig4}c], therefore the coupling Hamiltonian reads:
\beq H_{\text{coupling}} = H_{\text{IGP-vac}} + H_{\text{vac-SC}}, \eeq
with \beq H_{\text{IGP-vac}} =  \sum_{\langle i,j \rangle,\sigma}^{\{i, j\} \in \{\text{IGP, } \text{vac}\}} \left( \tau_{\text{IGP-vac}} c^\dagger_{i\sigma} c_{j\sigma} + \text{h.c.}\right), \eeq and 
 \beq H_{\text{vac-SC}} = \sum_{\langle i,j \rangle,\sigma}^{\{i, j\} \in \{\text{BLG, } \text{vac}\}} \left(\tau_{\text{NS}} c^\dagger_{i\sigma} c_{j\sigma} + \text{h.c.}\right), \eeq
where $\tau_{\text{IGP-vac}}\leq t$ and $\tau_{NS}\leq t$ are the hopping amplitudes between sites at the IGP and thee vacuum edge and between the vacuum edge and the superconductor, respectively. Their magnitudes control the inter-edge coupling rate and transparency at the NS interfaces.
 
The $I_c(\Phi)$ shown in Fig. \ref{fig4}(d,e) is computed using Eqs. \eqref{Ic} and \eqref{Icmax} and the spectrum of Eq. \eqref{HTB}.

\bibliography{band_inversion_references}

\end{document}